\documentclass[%
 reprint,
 amsmath,amssymb,
 aps,
]{revtex4-1}

\usepackage{graphicx}
\usepackage{dcolumn}
\usepackage{bm}
\usepackage{graphicx}
\usepackage{amsmath}
\usepackage{amsthm}
\usepackage{amsfonts}
\usepackage{amssymb, latexsym}
\usepackage{float}
\usepackage{color}
\usepackage{color,soul}
\usepackage{epstopdf}

\usepackage{ulem}

\begin{document}

\title{Active coating of a water drop by an oil film using a MHz-frequency surface acoustic wave}




\author{Avital Reizman}
 \affiliation{Department of Chemical Engineering, Technion Israel Institute of Technology, Haifa,
Israel.}

\author{Amihai Horesh}%
\affiliation{ 
Institute of Agricultural Engineering, Agricultural Research Organization, Volcani Center, Israel.
}%

\author{Lou Kondic}
\affiliation{%
Applied Mathematics, New Jersey Institute of Technology, NA, United States.
}%
\author{Ofer Manor}
 \email{manoro@technion.ac.il}
 \affiliation{Department of Chemical Engineering, Technion Israel Institute of Technology, Haifa,
Israel.}







\begin{abstract}
We employ a millimeter-scale piezoelectric acoustic actuator, which generates MHz-frequency surface acoustic waves (SAWs) in a solid substrate, to actively coat a drop of water by a macroscopic film of silicon oil as a paradigm for a small scale and low power coating system. 
We build upon previous studies on SAW induced dynamic wetting of a solid substrate, also known as the acoustowetting phenomena, to actively drive a model low surface-energy liquid -- silicon oil -- coat a model liquid object -- a sessile drop of water. The oil film spreads along the path of the propagating SAW and comes in contact with the drop, which is placed in its path. The intensity of the SAW determines the rate and the extent to which a macroscopically thick film of oil will climb over the drop to partially or fully cover its surface. The dynamic wetting of the drop by the oil film is governed by a balance between acoustic, capillary, and gravitational contributions. Introducing a water drop as an object to be coated indicates the opportunity to coat liquid phase objects by employing SAWs and demonstrates that oil films which are actuated by SAWs may traverse curved objects and liquid surfaces.
\end{abstract}

\maketitle

\begin{figure*}
\includegraphics[width=0.8\linewidth]{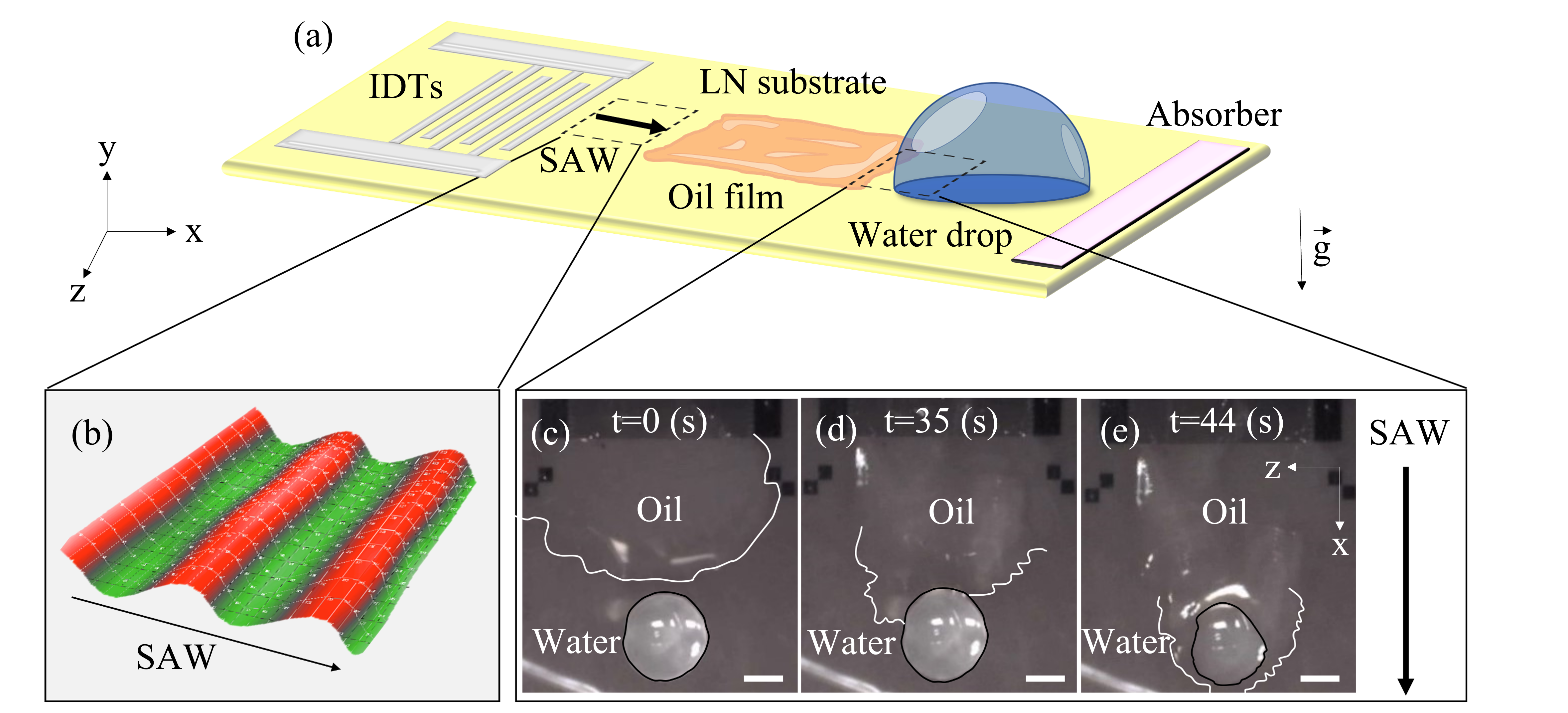}
\caption{A side view illustration of the experiment, where (a) we place a silicon oil film and a water drop atop a horizontal surface acoustic wave (SAW) actuator, comprising a piezoelectric lithium niobate (LN) substrate patterned with interdigitated electrodes (IDTs) and a SAW absorber (on the far side of the actuator) to eliminate reflections. (b) The SAW, propagating in the $x$ direction, generates spatiotemporal deformations in the solid surface in the $z$ direction, where red (green) illustrates the deformations above (below) the surface plane at rest, given at a specific time. (c) The experimental snapshots present the dynamic wetting of the solid substrate and the water drop by the oil film under SAW excitation (view from above).}
\label{fig:sys}
\end{figure*}

\section{Introduction}\label{sec1}

\par
Millimeter-scale piezoelectric acoustic actuators of MHz-frequency surface acoustic waves (SAWs) have increased in popularity in previous years for manipulating micro to nano-liter liquids and particulates therein.~\citep{miansari2016acoustic,pothuri2019rapid,horesh2017enhanced,horesh2020stabilizing} The actuators are piezoelectric substrates specifically designed to generate SAWs upon the application of a sinusoidal electrical signal,  which supports electromechanical resonance in the substrate. The piezoelectric effect in the substrate of the actuator linearly transforms the electrical signal to a nanoscale oscillating mechanical vibration of maximum amplitude near the surface of the devices -- SAW.~\citep{filipiak2021surface} SAW actuators are scalable in size, may be powered by a pocket size source, and are devoid of moving parts,~\citep{liu2022acoustofluidic,zhang2021manipulation,zhang2021powerful,shilton2015rapid} and hence are attractive for introducing low power and portable strategies for SAW actuated unit operations. Lithium-niobate (LN) based actuators,\citep{white1965direct} known for their high pyroelectricity and chemically inert surface, have been showing great promise for microfluidics and small scale manufacturing. These actuators are well known to manipulate and control millimeter to micrometer thick layers of liquid. In particular they support the dynamic wetting (coating) of the actuator substrate by oil \citep{rezk2012unique,rezk2014double} or water solutions.~\citep{altshuler2015spreading,altshuler2016free} However, using the actuators for coating objects, the spreading of liquid films by the SAW must endure geometrically structured objects. We demonstrate the application of a LN based SAW actuator for the active coating -- encapsulating -- of a model liquid object -- a drop of water -- by a macroscopic layer of a model coating liquid -- silicon oil -- as a paradigm for using a SAW actuator for coating objects.
\par

Coating a substrate, whether solid or liquid, is a fundamental process in the manufacturing industry. It is an endemic process in the production of automotive, electronic, medical and aerospace components alongside many others, traversing the manufacturing of aeroplanes and satellites down to the manufacturing of humble pens and pencils \citep{fotovvati2019coating}. Among the coating techniques found in industry one may count physical vapor deposition \citep{hassan2018antimicrobial,abbas2019fabrication} in which the coating material is evaporated from a solid or liquid source and then delivered as a vapor onto a substrate, where it condenses; chemical vapor deposition \citep{mittal2021nanofabrication,hong2020chemical}, where the coating of the substrate is via a chemical surface reaction with a vapor phase; electrodeposition \citep{walsh2020electrodeposition,szeptycka2016electrodeposition,pu2020electrodeposition}, where electrical potential is employed for the attraction of the coating materials in the form of ions or particulates to a charged substrate or via an electrochemical reaction at the surface of a charged substrate in contact with a precursor fluid; Sol-gel coating \citep{KLEIN198045}, which is a wet-chemical process for introducing a glassy or ceramic coating; spray coating \citep{tejero2019beyond,wang2021spray,WANG2010127,coatings1010017}, where a fast-moving flow of compressed air is steered into a stream of liquid coating and serves as a vector for transporting the coating to the target surface; and dip coating \citep{salles2019electrochromic,singh2020functionalization,byrnes2017pipeline}, where a coating liquid is put in direct contact with the target surface as in the happy blade approach or by dipping the target surface in a reservoir of a coating liquid. Dip coating as well as spray coating techniques are further employed to coat substrates by multiple immiscible liquid films \citep{belden_multilayer_2017,derflinger_multilayer_2009,grunlan_multilayer_2012}. In particular, coating of one liquid film by another is an intriguing problem when accounting for the dynamic and often unstable nature of the interface between the two coating liquid films \citep{horiuchi_simultaneous_2015,larsson_nonuniformities_2021}.\par

Lately, it was demonstrated that SAWs may be employed for precise and low power coating of small surfaces by oil films as a paradigm for coating objects of lateral length scales of millimeters to centimeters.~\citep{rezk2012unique,rezk2014double} In  particular, vibration in a substrate in the form of a surface acoustic wave (SAW), undergoing frequencies comparable to the HF and VHF radio frequency range, may invoke the continuous spreading of sub micrometer to hundreds of micrometer thick films of a fully wetting silicon oil along or opposite the path of the propagating SAW. Partially wetting water films under similar excitation were also found to undergo similar coating flows.~\citep{altshuler2015spreading,altshuler2016free} However, in the latter case, capillary stresses had to be reduced to a level below that of the acoustic stresses in the film to render dynamic wetting; otherwise capillary stresses arrest the coating flow. \par
There are several mechanisms by which SAW actuates the dynamic wetting of substrates by liquid films.~\citep{manor_dynamics_2015,morozov_vibration-driven_2018} Briefly, A SAW in a solid substrate leaks ultrasonic waves of the same frequency and of a wavelength $\lambda_l$ into the nearby fluid. The ratio between $\lambda_l$ and the thickness of the film \citep{shiokawa_liquid_1989} determines the specific mechanism that governs film dynamics. In previous studies on micron and tens of micron thick liquid films and SAW frequency of $20$-$40$~MHz, \citep{rezk2012unique,rezk2014double,manor_dynamics_2015,morozov_vibration-driven_2018}
the thicknesses of the film were smaller than $\lambda_l$. The governing mechanism for the actuation of film dynamics was found to be the Stokes drift \citep{Schlichting:1932p447, Rayleigh} of liquid mass due to the convection of momentum in an acoustic boundary layer flow near the solid substrate.  

\par
We consider oil films that are hundreds of micron thick and a SAW frequency of 20 MHz. In this case, the thickness of the film is greater than the wavelength of the acoustic leakage ($\lambda_l\approx 80~\mu$m). The dominant acoustic streaming mechanism is Eckart Streaming. Flow is predominantly generated by the attenuation of the SAW in the actuator \citep{shiokawa_liquid_1989}. The Attenuation of SAW spatially alters the intensity of the acoustic leakage in the liquid and hence generates Reynolds stress variations, i.e., net body force therein. The Reynolds stress in the liquid film brings about the spreading of the film on the surface of the actuator and hence the dynamic wetting of the actuator along the path of the SAW.  

\par
The capability of a liquid film, actuated by a SAW, to dynamically wet a solid substrate and to overcome a textured substrate or coat a solid or liquid object atop the SAW actuator was not studied to date, albeit it was used for SAW based PCR systems on a chip \cite{B412712A}, where it appears that the coating of small water drops by oil was governed by capillary effects. Here, we consider the problem of using a SAW to actuate the coating of a liquid object by a low surface energy coating. For this endeavor, we employ a model experimental system in which we employ a SAW actuator to coat a water drop by a macroscopic layer of silicon oil. The water drop is a model liquid obstacle, which we place in the path of the silicon oil film. The film dynamically wets the solid substrate under the action of the propagating SAW therein and partially or fully climbs over the drop upon contact. 

Our goal is to realize the physical conditions that support the active coating of a water drop by a macroscopic layer of oil film. We briefly describe our experimental procedure in Experiment, give our results alongside a dedicated discussion in Results and Discussion, and summarize and conclude our work in Summary and Conclusion.

\begin{figure*}
\includegraphics[width=0.8\linewidth]{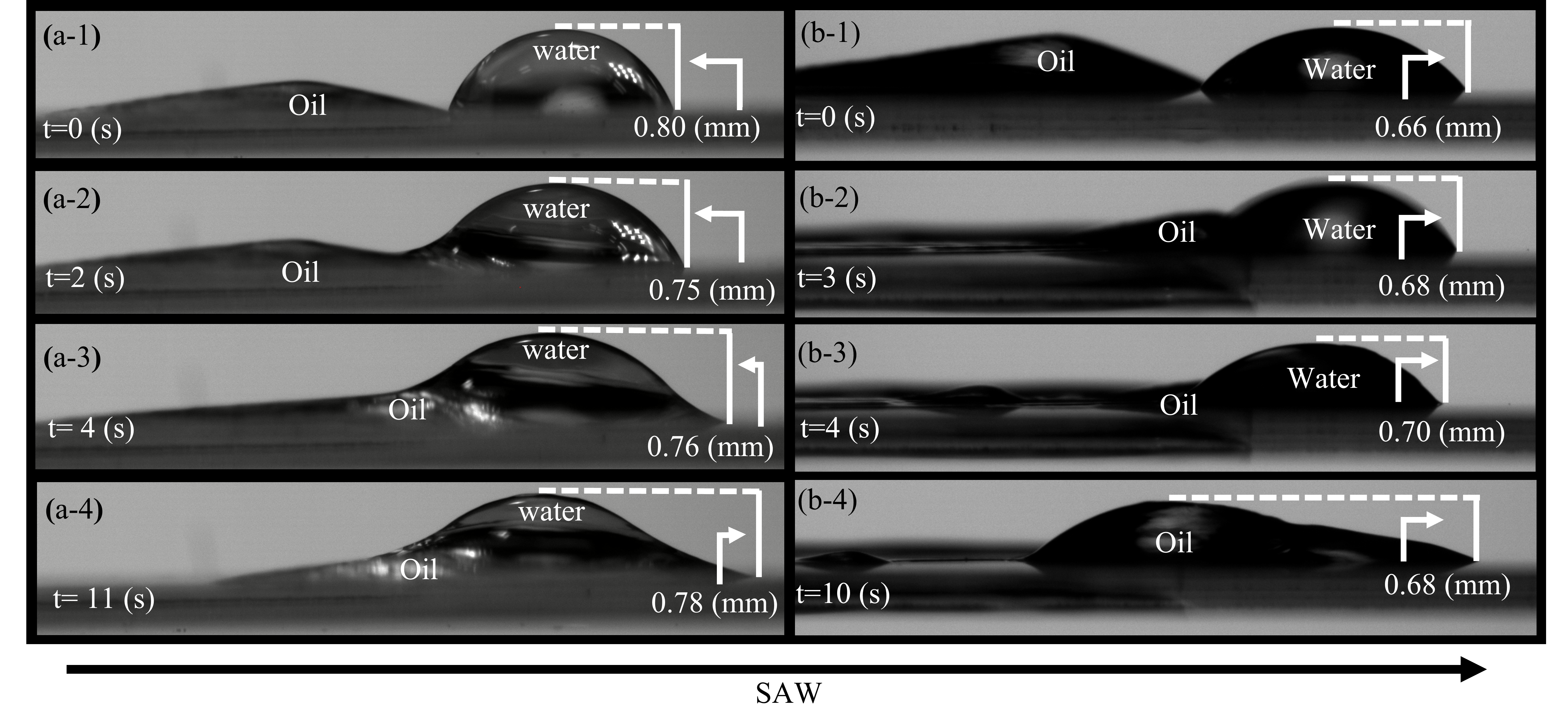}
\caption{Time-lapse of (side view) images, which portray the dynamic wetting of a sessile drop (right) by a silicone oil film (left) under SAW intensities of (a) SDA$\sim4.0$~\AA, where the oil film partially coat the drop, and (b) SDA$\sim$ $9.3$~\AA, where the oil film fully coat the drop. Initially (t=$0$~s), we show the moment of contact between oil and water; at following times, we demonstrate the coating of the water drop by the oil film which spreads to the right. {The light exposure is different between the sections.}}
\label{fig:1Volts_side}
\end{figure*}
\begin{figure}
\center
\includegraphics[width=\linewidth]{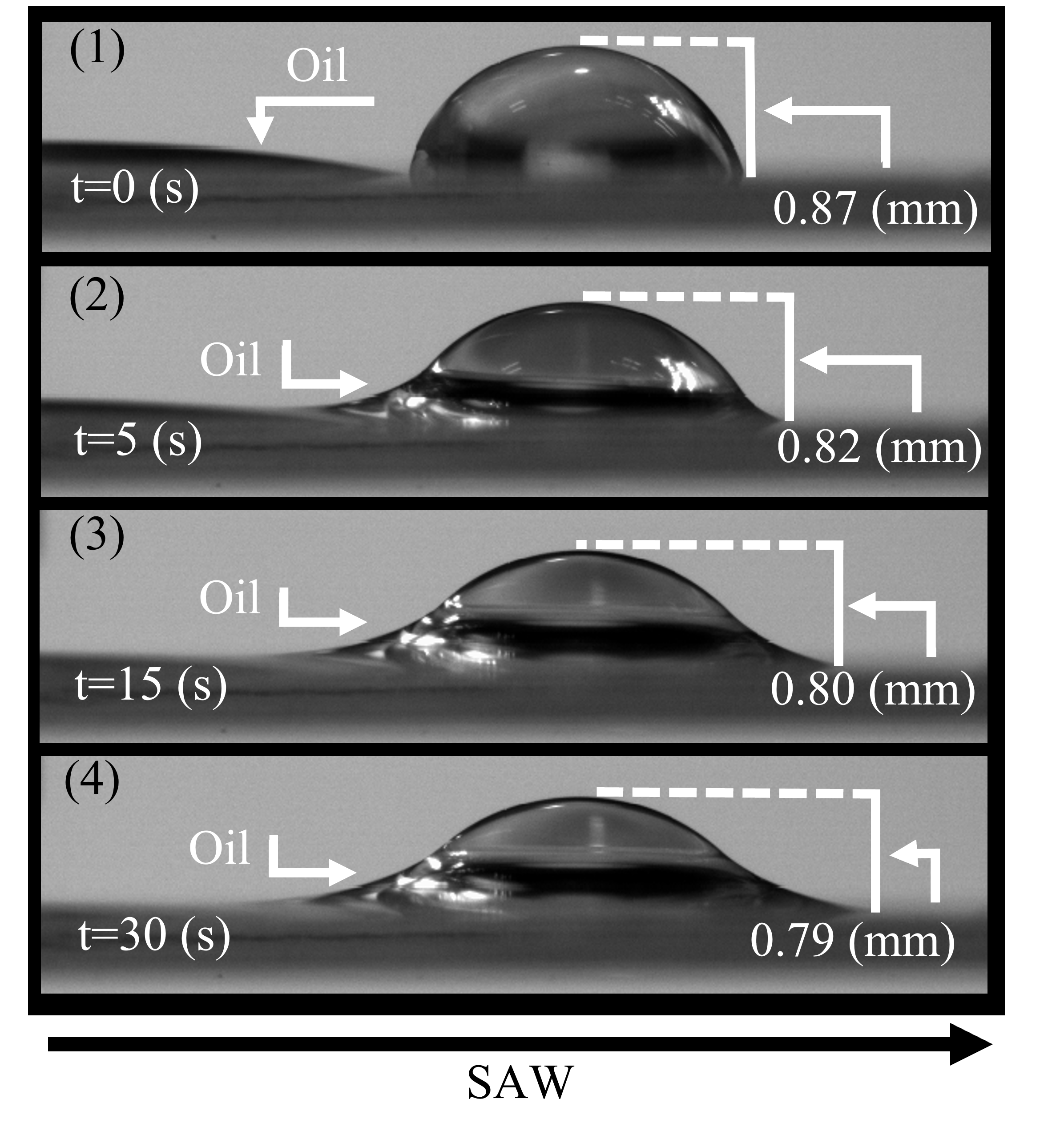}
\caption{{Time-lapse of (side view) images, which portray the dynamic wetting of a sessile drop (right) by a silicone oil film (left) in the absence of an acoustic wave. Initially (t=$0$~s), we show the moment of contact between oil and water; at following times, we demonstrate the partial wetting of the water drop by the oil film, where the system appears to obtain steady state at $t=30$s.}
}
\label{fig:0Volts_side}
\end{figure}

\section{Experiment}\label{sec4}

\par
We fabricated an actuator -- a SAW device -- by using photolithography to place interdigiated electrodes (IDTs), comprising $5$ nm titanium as adhesion layer and $1$~$\mu$m aluminum, atop a piezoelectric 128$^\circ$ y-cut lithium niobate ($\text{LiNbO}_{\text{3}}$) substrate. The actuator generates a $20$~MHz-frequency Rayleigh type SAW in the solid substrate.

Fig~\ref{fig:sys} illustrates the use of a SAW actuator for coating a water drop: We initially placed A $3.0$ $\pm$ $0.05$ ~$\mu$L sessile drop of deionized (DI) water and a puddle of a 100~cSt viscous silicone oil (SIGMA-ALDRICH, Merck) atop a SAW actuator. Upon the application of power to the actuator, the SAW of a wavelength of $200~\mu$m propagates at a phase velocity of \citep{kushibiki1999accurate} 3880~m/s away from the electrodes near the upper surface of the actuator (which faces the two liquids). Firstly encountering the silicon oil film, the SAW leak ultrasonic waves of a wavelength of approximately $\lambda_l\approx80~\mu$m into the oil phase. The ultrasonic waves generate Reynolds stress and Eckart streaming in the $100-200~\mu$m thick oil film of a nominal dynamic viscosity of 100~cSt and nominal surface tension of 20~mN/m at room temperature. The SAW actuates the dynamic wetting of the substrate by the oil film toward the drop. As the actuated oil film comes in contact with the sessile drop, which is pinned to the solid substrate by capillary stresses~\citep{altshuler2015spreading}, it climbs over the drop to partially or fully coat its surface. The level to which the oil film climbs over the drop is found dependent on the normal surface displacement amplitude of the SAW (SDA). We give in Supporting Information graphs which connect the SAW to the voltage, which we generate using our signal generator (R\&s SMB100A microwave signal generator) and apply to an 5X amplifier (Model A10150, Tabor Electronics Ltd.) and then to the SAW devices. Moreover, To capture the film dynamics, We used a side-view camera (Data Physics; OCA 15Pro) and a top-view USB camera (Dino-Lite).


\par
The SAW imposes comparable lateral and transverse harmonic displacement in time and space in the LN substrate. The displacement amplitude is proportional to the voltage, i.e., to the square root of the power applied to the actuator. We measured the (normal) surface displacement amplitude (SDA) of the SAW at the spatial region of interest on the LN substrate; specifically, we measured the displacement at the initial location between the oil and water drop for each experiment. For the measurement, we used a scanning laser Doppler vibrometer (MSA-500, Polytech) for each input voltage level and verified that the SAW is a propagating wave. To avoid the presence of a standing SAW on the actuators, we placed a layer of wet paper (Kimberly-Clark Professional : KIMWIPES Ex-L) at the far side of the actuators (opposite side to the IDTs) as wave absorbers. Moreover, in a bare SAW device, the attenuation length of the $20$ MHz-frequency SAW is approximately $1$ meter. This translates to negligible attenuation on our device, which is a couple of centimeters long. Under the oil film and water drop, the SAW should attenuate appreciably however to lesser extent than the attenuation level of the SAW under half-space of water or oil. In this case the attenuation length of the 20 MHz-frequency SAW is approximately 1 millimeter.~\citep{doi:10.1063/1.1754894} Hence, the actual position of the drop on the SAW device may introduce a change to the height in which the oil film climbs atop the drop. {To reduce variability in the experiment, we kept the position of the drop and oil film on the SAW device the same during our different experiments, i.e., 3 mm away from the IDT.}

Prior to each experiment, we rinsed the actuator under toluene, acetone, 2-propanol, ethanol (Bio-Lab Ltd.) for $20$~s and finally under HPLC water (HPLC plus, Sigma-Aldrich). We then used Pogo pins (BC201403AD, Interconnect Devices, INC.) to connect the IDTs on the actuator to a signal generator (R\&S SMB100A) and a signal amplifier (A10150, Tabor Electronics Ltd.). The SAW was generated by imposing different levels of a continuous sinusoidal input voltage at a fixed frequency of $20$ MHz. {Moreover, initially, we examined the rate of water evaporation from the drop in the absence of silicon oil by monitoring the drop height over the time of the experiments which later on included oil. The change in the drop height in these measurements was found negligible. We then conducted a parametric study of the coating of water drops by silicon oil atop the SAW actuator.}

\section{Results and Discussion}\label{sec2}
We introduce representative time lapse images showing the partial and full coating of a water drop by oil film in Fig~\ref{fig:1Volts_side}. The figure shows a side view of the experiment for two representative cases. Fig~\ref{fig:1Volts_side} (a) illustrates the partial coating dynamics of a water drop by oil film for a SAW normal displacement amplitude at the surface of the actuator (SDA) of $\sim$ $7.0$ $\pm$ 0.59 ~\AA, and Fig.~\ref{fig:1Volts_side} (b) illustrates the full coating dynamics for an SDA of $\sim$ $9.3$ $\pm$ $0.59$ ~\AA. In both cases, we show the initially separated oil film (left) and water drop (right), and then time laps of the climb of the macroscopic oil film over the drop. The oil film partially coat the water drop when actuated by SAW of the lower SDA value, leaving the upper part of the drop devoid of a macroscopic oil film (although it may be coated by a molecular layer of oil due to capillary stresses). The oil film fully coats the water drop by a layer of a macroscopic thickness when actuated by a SAW of the greater SDA value. The climb height of the oil over the drop is known to be the result of acoustic, capillary, and gravitational stresses in the therein.~\citep{horesh2019acoustogravitational} Corresponding movies of the experiment are given in Supporting Information. 

\begin{figure}
\center
\includegraphics[width=\linewidth]{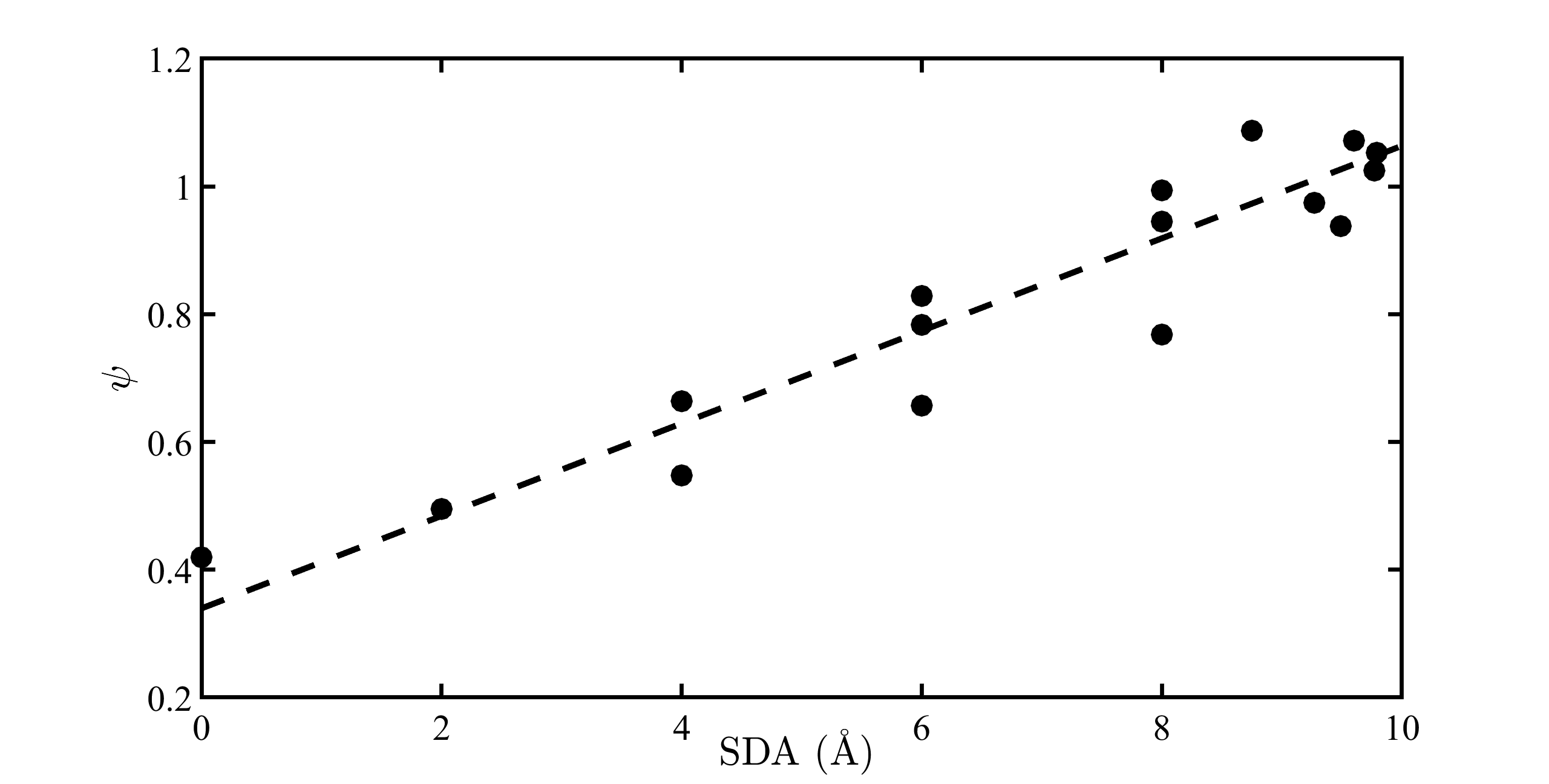}
\caption{SAW intensity (SDA) variations of the oil film height ($h_c$) to drop height ($h_i$) ratio, $\psi\equiv h_c/h_i$, which indicates the level of climb of the oil film over the drop, showing a approximately linear trend. Note that for  $\psi<1$, the oil does not reach the top of the drop, and for $\psi>1$, the oil film extends over the top of the drop, fully coating the latter; the line is a guide to the eye. {The characteristic errors are: $\psi= \pm$ $0.04$ and SDA = $\pm$ $0.73$ ~\AA.} }
\label{fig:PhiPsi}
\end{figure}

To evaluate the contribution of the SAW to the coating process, we further give time lapse images of the interaction of the oil film and a water drop on our actuator, but in the absence of SAW (no power is applied to the actuator) in Fig~\ref{fig:0Volts_side}. The figure demonstrates that in the absence of SAW, the oil slowly and spontaneously spreads across the LN surface due to its low surface tension. Once encountering the water drop, the macroscopic oil climbs to a finite height, partially coating the drop. This is quantified in Fig~\ref{fig:PhiPsi}, where the oil film climbs over the drops in our experiment to $\psi$=$0.42$ $\pm$ $0.04$ of the drop heights of $h_i=1.05$ $\pm$ $0.02$ mm in our experiments. The climb height in this case is associated with a balance between capillary and gravitational stresses in the oil film;\citep{xue2020marangoni} here $\psi\equiv h_\text{c}/h_\text{i}$ represents the ratio between the measured height to which the oil film climbs over the water drop, $h_\text{c}$, and the height of the water drop prior to SAW excitation $h_\text{i}$. The latter appears to remain the same throughout the experiment. As a note, in our experiments, the volume of the drops is $3.00\pm $0.05$~\mu$L and the height of the water drops is $h_{i}$= $0.88$ $\pm$ $0.02$ mm. {The main reason for variations in drop heights between different experiments is most likely the three phase contact angle hysteresis. The hysteresis renders the height of a drop dependent on the specifics of placing the latter on the actuator surface.} ~\citep{mittal_guide_2009} Increasing the applied power to the SAW, leads to an increase in $\psi$. SAW induced Reynolds stress in the oil film modifies the previously mentioned equilibrium and supports a greater climb height of the oil film.~\citep{horesh2019acoustogravitational} Above an SDA threshold of approximately $\sim$ $8.0$~\AA, we observe that the macroscopic oil film fully covers the drop in most experiments. 

\begin{figure}
\center
\includegraphics[width=\linewidth]{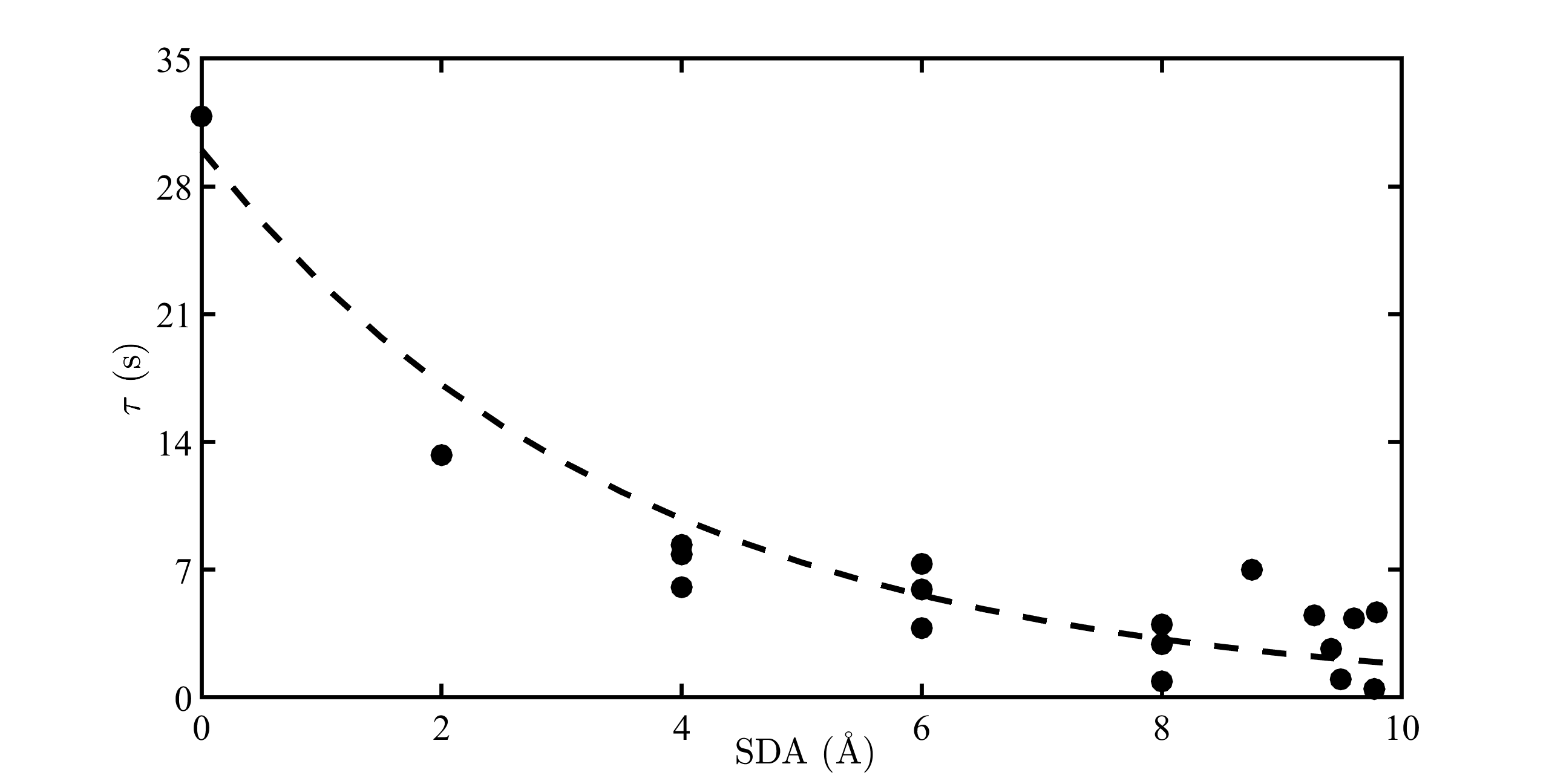}
\caption{We show the time to reach steady state ($\tau$) against the intensity of the SAW. From the moment the oil film touches the drop to the moment the oil film attains a steady state height for SDA$<\sim$ $8.0$~\AA, or reaches the top of the drop for SDA$>\sim$ $8.0$~\AA. A decaying trend is observed. the line is a guide to the eye. {The characteristic errors are: $\tau= \pm$ $0.02$ s and SDA = $\pm$ $0.73$ ~\AA.}} 
\label{fig:Coating_time_vs_dis}
\end{figure}

\par
To give an indication of the dynamics of the coating process, we show in Fig.~\ref{fig:Coating_time_vs_dis} the influence of the SAW intensity (SDA) on the time, $\tau$, in which the oil film climbs over the drop. We define $\tau$ as the time that passes from the moment of contact between the oil film and the drop to the time in which the oil film reaches a steady state height (when it does not reach the top of the drop) or to the moment it reaches the full height of the drop. In particular, a drop may be partially or fully coated by the oil film at time $\tau$. The system reaches a steady state when the oil film partially coats the water drop for approximately SDA$<8.0$~\AA. The oil film fully coat the drop for approximately SDA$>8.0$~\AA. 

\begin{figure}
\center
\includegraphics[width=\linewidth]{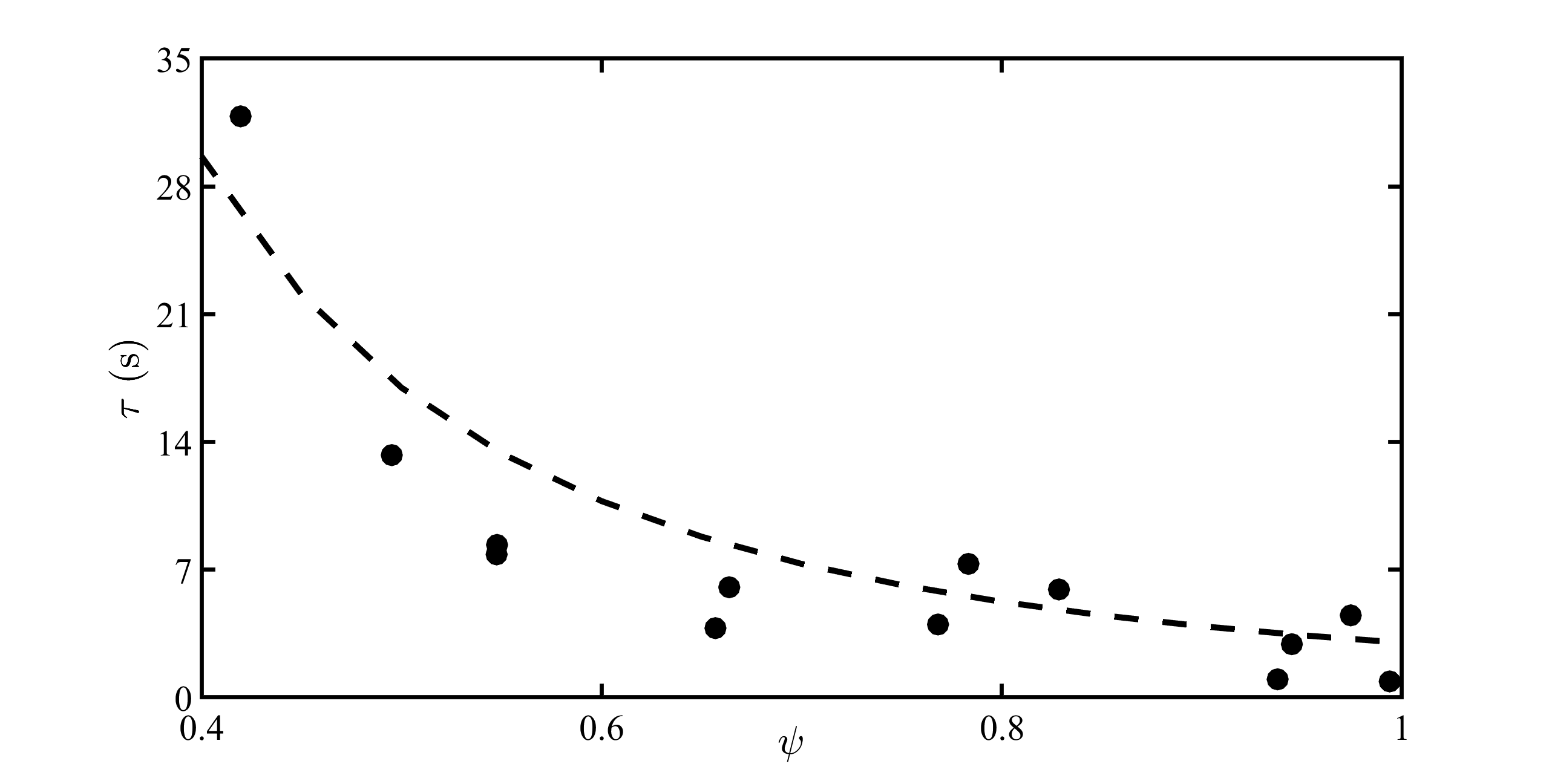}
\caption{ We show the time to reach steady state ($\tau$) against the ratio between the climb height of oil film and the original height of the water drop. {Each point on the graph represents a different experiment at different intensity of the SAW as follows (from left to right): $0.0$, $2.0$, $4.0$, $4.0$, $6.0$, $4.0$, $8.0$, $6.0$, $6.0$, $9.5$, $8.0$, $9.3$, $8.0$ ~\AA }. Relative height of oil climb, $\psi$, variations of the coating time, $\tau$, indicating an inverse relationship, which is mediated through the magnitude of the SAW intensity (SDA), where the line is a guide to the eye. {The characteristic errors are: $\psi= \pm 0.04$ and $\tau = \pm 0.02$ s.}
} 
\label{fig:Coating_time_vs_Psi}
\end{figure}

 In the figure, we demonstrate the slow climb of the oil film atop the drop in the absence of the SAW (SDA=0) and the remarkable reduction of the value of $\tau$ in the presence of the SAW (SDA$>0$). Moreover, in Fig.~\ref{fig:Coating_time_vs_Psi}, we connect $\tau$ and the steady state height of climb of the oil film, $\psi$. The figure indicates a reciprocal connection between the two quantities. That is, increasing the acoustic intensity (SDA) results in shorter time of climb to a greater height of the drop. 

\section{Summary and Conclusion}\label{sec3}
Here we demonstrate the application of a millimeter size LN based acoustic actuator, which generates a MHz-frequency SAW, to support the active coating of a sessile water drop by a silicone oil film. The difference in the surface energy of the two liquids results in different responses of the two liquids to the SAW. The low surface energy oil dynamically wets the solid substrate and the water drop. The high surface energy water drop keeps sessile. Hence the oil and water system makes a well defined and ideal case to explore the SAW induced dynamic wetting of a non planar object and in particular a liquid object.

In the absence of SAWs, the oil dynamically wets the LN surface. Upon contact between the oil film and the water drop, the macroscopic oil film partially climbs atop the drop as a result of a gravitational-capillary stress balance. Gravitational and capillary contributions to the climb arrest the motion of the macroscopic oil film when it reaches a partial height of the drops in our experiments. 

Upon introducing a propagating SAW in the LN substrate, the oil film dynamically wets the substrate along the path of the SAW. Once the oil film makes contact with the drop, it climbs over the surface of the latter. The intensity (SDA) of the SAW determines the extent of the drop coverage by oil. Greater levels of SDA results in greater coverage of the drop surface. Above an SDA threshold of $8.0$~\AA, we observe that the oil film fully coat the drops in our experiments. 



\section{Acknowledgement}\label{sec5}
We are grateful for the support of this study by the Israel Science Foundation, grant number $441$/$20$ and by the United States Israel Binational Science Foundation, grant number $2020174$.

\bibliographystyle{elsarticle-num}
\bibliography{references}

\end{document}